\documentclass[ reprint,amsmath,amssymb,aps,pra,]{revtex4-2}
\usepackage{graphicx}
\usepackage{dcolumn}
\usepackage{bm}
\usepackage{hyperref}
\usepackage[utf8]{inputenc}
\usepackage{xcolor}
\hypersetup{
	colorlinks=true,
	linkcolor=blue,
	citecolor=blue,
}
\begin{document}
\preprint{APS/123-QED}
\title{Thermoelectrical potential and derivation of Kelvin relation for thermoelectric materials}
\author{Si-Kun Chen}
\author{Hong-Xin Zhu}
\author{Hai-Dong Wang}
 \email{hdwang@tsinghua.edu.cn}
\author{Zeng-Yuan Guo}
 \email{demgzy@tsinghua.edu.cn}
\affiliation{Key Laboratory for Thermal Science and Power Engineering of Ministry of Education, Department of Engineering Mechanics, Tsinghua University, Beijing 100084, China.}
\date{\today}
\begin{abstract}
		Current research on thermoelectricity is primarily focused on the exploration of materials with enhanced performance, resulting in a lack of fundamental understanding of the thermoelectric effect. Such circumstance is not conducive to the further improvement of the efficiency of thermoelectric conversion. Moreover, available physical images of the derivation of the Kelvin relations are ambiguous. Derivation processes are complex and need a deeper understanding of thermoelectric conversion phenomena. In this paper, a new physical quantity “thermoelectrical potential” from the physical nature of the thermoelectric conversion is proposed. The quantity is expressed as the product of the Seebeck coefficient and the absolute temperature, i.e., ST. Based on the thermoelectrical potential, we clarify the conversion of the various forms of energy in the thermoelectric effect by presenting a clear physical picture. Results from the analysis of the physical mechanism of the Seebeck effect indicate that the thermoelectrical potential, rather than the temperature gradient field, exerts a force on the charge carriers in the thermoelectric material. Based on thermoelectric potential, the Peltier effects at different material interfaces can be macroscopically described. The Kelvin relation is rederived using the proposed quantity, which simplified the derivation process and elucidated the physical picture of the thermoelectrical conversion.
\end{abstract}
\maketitle
\section{\label{sec:level1}INTRODUCTION}
		The study of thermoelectricity is of great significance in basic science and engineering application. As early as 1821, Seebeck \cite{seebeck1822ueber} experimentally observed for the first time the thermoelectric effect in materials. He discovered that a magnetic compass placed near two thermocouple connections would be deflected when they were at different temperatures, opening the door of thermoelectric research. In 1834, Peltier \cite{peltier1834nouvelles} observed a contrasting phenomenon, i.e., when an electric current passes through an isothermal connection of two different metals, the connection point absorbs or releases heat depending on the direction of the electric current. These two phenomena are now well-known as the Seebeck and Peltier effects, respectively \cite{beretta2019thermoelectrics,3aec202aacb94720a87877af45bbcb80}. In 1851, Lord Kelvin \cite{thomson18574} proposed a third thermoelectric effect (now known as the Thomson effect), i.e., when an electric current is passed through a homogeneous conductor in which a temperature gradient exists, heat is absorbed or emitted in the conductor, in addition to the Joule heat associated with electric resistance. Kelvin further investigated the relationship between these three effects by using equilibrium thermodynamics and proposed the famous Kelvin's first relation: 
\begin{eqnarray}
	\mathit{\Pi}=-ST
\end{eqnarray}
where $\mathit{\Pi}$ , \textit{S} and \textit{T} are the Peltier coefficient, Seebeck coefficient and temperature, respectively. In the 1950s, Ioffe \textit{et al}. \cite{ioffe1958semiconductor} proposed the use of semiconductors as thermoelectric materials, which renewed the interest of researchers and marked a new milestone in the development of thermoelectric materials. They also proposed the concept of figure of merit to evaluate the thermoelectric performance of the materials, i.e.,$\mathit{Z}=\mathit{S^2}\mathit{\sigma}/\mathit{\lambda}$. In the following decades, conventional thermoelectric materials based on semiconductor systems, such as $\mathrm{Bi_{2}Te_{3}}$ \cite{hong2016n}, PbTe \cite{jia2022realizing} and SiGe \cite{joshi2008enhanced}, have been remarkably developed. In addition, organic polymer materials, crystalline porous organic materials and Mxene have also been gradually formulated as thermoelectric materials. Jing feng \textit{et al}. \cite{li2023wide} obtained the Nb composite \textit{n}-type $\mathrm{Mg_{3}(Sb, Bi)_{2}}$ materials with \textit{ZT} values up to 2.04 by inlaying highly conductive metal nanomaterials of Nb in the grain boundaries. Superlattice materials have been demonstrated to possess high thermoelectric properties, with values reaching 1.7 \cite{su2022high}. 
\par
Current study on thermoelectricity is primarily focused on the exploration of materials with higher ZT values. These materials can be enhanced by the tuning of the transport properties of electron and phonon in the material, such as carrier concentration, mobility, bandgap width and the phonon and electron mean free paths. Engineering can be specifically performed by improving the synthesis \cite{tsuchikawa2020unique}, doping \cite{liang2016large} and formation of composites \cite{snyder2008complex} to enhance the electrical conductivity and Seebeck coefficient of the material, whilst decreasing its thermal conductivity to improve their thermoelectric properties. However, a deeper and more fundamental understanding of the thermoelectric phenomena is lacking. This limitation is not conducive to the further improvement of the efficiency of thermoelectric conversion. For example, the Seebeck effect is prone to misinterpretation as the balance of the temperature difference with the electrical potential difference. Moreover, only a microscopic analysis but no macroscopic description of the Peltier effect at interfaces has been performed. As a consequence, the physical image of the derivation of Kelvin relation in the existing literature is ambiguous. The derivation is complex and lacks an in-depth understanding of the physical nature. In the current paper, the “thermoelectrical potential” is proposed as a new physical quantity, resulting the clarification of the conversion between different forms of energy. Thus, a more well-defined physical picture is obtained, simplifying the derivation of the Kelvin relation. 
\section{\label{sec:level2}THE THERMOELECTRICAL POTENTIAL- A NEW PHYSICAL QUANTITY FOR THERMOELECTRIC CONVERSION PROCESSES}
\subsection{\label{sec:level3}Why do we introduce the thermoelectrical potential?}
(1) The Seebeck coefficient is defined as $\mathit{S}=\Delta\mathit{V}/\Delta\mathit{T}$. This parameter is a measure of the ability of a material to convert thermal energy into electrical energy over a temperature gradient. In theory, at this point, a balance exists between the temperature difference $\Delta\mathit{T}$ and the potential difference $\Delta\mathit{V}$, esulting in the absence of current inside the component. However, at this point, the electric field strength caused by the voltage difference $\Delta\mathit{V}$ s not zero, but the thermal or temperature field in the thermoelectric element itself cannot generate a force to balance the electric field force. Therefore, for an open-circuit state, another potential difference and another corresponding electric field must exist in the thermoelectric element. In such case, this “another” electric field is equal in strength but opposite to the electric field caused by the voltage difference, resulting in the absence of an electric current inside the element. The micro-mechanism of another electric field is described as when the temperature of one end of the thermoelectric element increases, the holes at the high-temperature end of the \textit{p}-type element have higher kinetic energy compared to those at the low-temperature region. In addition, more carriers are excited into the conduction or valence bands, and the concentration of charge carriers is higher, leading to the diffusion of holes from the high-temperature end to the low-temperature one. The uneven distribution of holes in the thermoelectric element results in an electrical potential difference between the two ends of the thermoelectric element. This condition inversely prevents the holes at the high-temperature end from moving to the low-temperature one.\par
(2) The Peltier coefficient is defined as $\mathit{\Pi}=-\mathit{Q}/\mathit{I}$, where \textit{Q} represents the heat flow at the interface, and \textit{I} represents the flux of charge carriers at the interface. Here, the Peltier coefficient only reflects the ratio of two physical quantities, namely, the heat flow and electric current, at the interface. The quantity $\mathit{\Pi}=-\mathit{Q}/\mathit{I}$ does not explain the physical mechanism of the energy conversion between two different energy forms. The physical mechanism of the Peltier effect at the microscopic level has been reported. The \textit{n}-type and \textit{p}-type materials have distinct energy levels of charge carriers. At the interface between these materials, electrons will transition to higher energy levels and absorb heat from the outside when they move directionally from the \textit{p}-type to the \textit{n}-type materials. When the direction of the current is reversed, the electrons move in the reverse direction, transition to lower energy levels and release heat to the outside \cite{nikam2014review,radermacher2007integrating}. However, the heat conduction, electrical charge conduction and their conversion processes are commonly investigated using macroscopic physical quantities, such as voltage, current, temperature, Seebeck and Peltier coefficients. Analysing the energy levels of the charge carriers is important to elucidating the micro-mechanism of the Peltier effect. However, corresponding macroscopic physical quantities are unavailable when the thermoelectric processes at the interface are analysed. In particular, when the current passes through the interface between two different materials, the voltage \textit{V} on both sides of the interface does not change and the electrical energy flow \textit{IV} remains constant. These phenomena indicate that another potential difference must exist at the interface, and when the current flows through the interface, another energy flow difference converted from the heat flow exists. We propose the macroscopic potential corresponding to the energy levels of the charge carriers as the “thermoelectrical potential” and the energy flow through the interface as the “thermoelectrical energy flow”. This definition indicates that the heat absorption or release at the interface due to Peltier effect is not the conversion between thermal and electrical energy flows, as described in the existing literatures, but the conversion of thermal energy flow to another kind of energy flow.
\subsection{\label{sec:level4}How to introduce the thermoelectrical potential?}
(1) From the analysis of the physical mechanism of the Seebeck effect in section \ref{sec:level3}, in the open-circuit state, another electric field exists in the material and exerts a force on the free charge in the direction opposite to that of the strength of the electrostatic field. After being balanced with the electrostatic field, the current inside the material becomes zero. Thus, this potential should have the same unit as the electrical potential. 
\par
The Seebeck coefficient, which is considered as a constant, is defined as follows:
\begin{eqnarray}
	\mathit{S}=-\dfrac{\Delta\mathit{V}}{\Delta\mathit{T}}
\end{eqnarray}
The equation can be rewritten by moving the temperature difference to the left side of the equal sign as follows:
\begin{eqnarray}\label{eq3}
	\mathit{S}(\mathit{T}_{1}-\mathit{T}_{2})=\mathit{S}\mathit{T}_{1}-\mathit{S}\mathit{T}_{2}=-\Delta\mathit{V},
	\mathit{T}_{1}>\mathit{T}_{2}
\end{eqnarray}
\par
Thus, the electrical potential difference is balanced by the difference in the physical quantity \textit{ST}, not the temperature difference. On the right side of the equation is the electrical potential difference between the two ends of the material, whilst that on the left side is the difference between the product of the Seebeck coefficient and the absolute temperature of the material at each end. As mentioned earlier, it represents another potential difference that is equal in strength and opposite in direction to the electric field.  Because the physical quantity \textit{ST} is a function of temperature and has the dimension of voltage, it is called thermoelectric potential.
\par
(2) The dimension of the physical quantity \textit{ST} is volt. Thus, to analyse \textit{ST} conveniently, the thermoelectrical potential \textit{ST} can be written as $\mathit{V_{t}}=\mathit{ST}$, and the Seebeck coefficient as follows:
\begin{eqnarray}
	\mathit{S}(\mathit{T}_{1}-\mathit{T}_{2})=\mathit{V_{t}}_{1}-\mathit{V_{t}}_{2},\mathit{T}_{1}>\mathit{T}_{2},\mathit{V_{t}}_{1}>\mathit{V_{t}}_{2}
\end{eqnarray}
\par
The absolute value of the thermoelectrical potential $\mathit{V_{t}}$ is not equal to the potential \textit{V} at this time. However, when $\mathit{T}_{2}$ is equal to zero, the absolute value of the thermoelectrical potential $\mathit{V_{t}}$ is equal to the electrical potential \textit{V}. 
\par
\begin{figure}[b]
	\includegraphics[width=8.6cm]{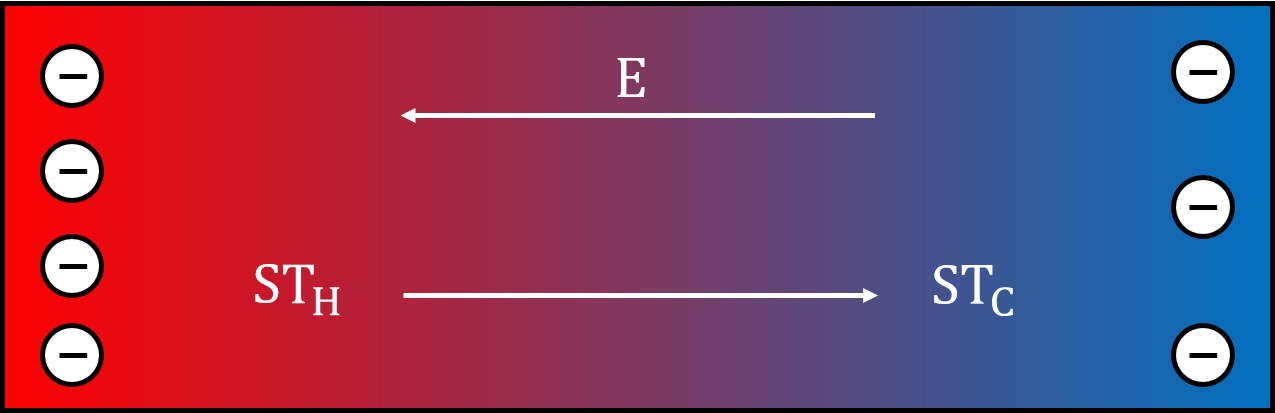}
	\caption{\label{fig:1}Thermoelectric potential difference within the same material in an open-circuit state.}
\end{figure}

For a \textit{p}-type semiconductor, at the higher temperature end, the charge carriers are excited by heating. Then, the thermoelectrical potential rises, forming a thermoelectrical potential field within the material. This phenomenon will drive the charge carriers to move from the high-temperature end to the low-temperature end, resulting in a charge build-up at the low-temperature end. In an open-circuit state, the reverse electrical potential field and the thermoelectrical potential field, which are equal in strength and opposite in direction due to the charge build-up caused by the carrier movement, are balanced, such that the total carrier mobility in the \textit{p}-type material and the total current are both zero. Combined with the rewritten Seebeck coefficient expression Equation \eqref{eq3}, $\mathit{S}\mathit{T}_{1}-\mathit{S}\mathit{T}_{2}$ is the thermoelectrical potential difference that balanced by the electrical potential difference (Fig.\ref{fig:1}).
\section{\label{sec:level5}DERIVATION OF KELVIN RELATION BASED ON THE THERMOELECTRICAL POTENTIALS}
\subsection{\label{sec:level6}Derivation of existing theories}
The Kelvin relation was first obtained by Kelvin, and the equilibrium thermodynamic method for deriving the Kelvin relation was also presented. He conceived a thermoelectric circuit consisting of two thermoelectric materials. Two junctions between the two materials are in contact with a heat source at different temperatures \cite{domenicali1954irreversible}. According to Kelvin, in a fully reversible closed loop, the heat absorbed at different temperatures obeys a linear equation with a coefficient that is the reciprocal of the temperature \cite{thomson1857ix}. For thermoelectric circuits in which any irreversible factor is not considered, the heat flows from the Peltier and Thomson effects cancel each other out \cite{lippmann1907analogie}. This phenomenon is the law of conservation of entropy for thermoelectric circuits. Moreover, in this thermoelectric circuit, the law of conservation of energy provides an energy relation that includes the Seebeck, Peltier and Thomson effects. Based on the above considerations, Kelvin first derived the famous Kelvin relation. Although Thomson gave the correct Kelvin relation, his proof was considered problematic. As Onsager \cite{onsager1931reciprocal} stated, “Thomson's relation has not been derived entirely from recognised fundamental principles, nor is it known exactly which general laws of molecular mechanics might be responsible for the success of Thomson's peculiar hypothesis.” In this historical context, Onsager proposed a profound symmetry relation, i.e., the famous Onsager reciprocal relation, between irreversible transfer processes. In 1948, Callen \cite{callen1948application} utilised the Onsager reciprocal relation to rederive the Kelvin relation. Since the Onsager relation is more in line with actual physical scenarios, Callen’s derivation is apparently more convincing than that of Thomson’s. However, the proof of the Onsager reciprocal relation needs to resort to the fluctuation dissipation theorem and the assumption of microscopic reversibility. Thus, the derivation process is relatively complex, resulting in the ambiguity of the macroscopic physical image of the thermoelectric conversion process. 
\subsection{\label{sec:level7}Derivation of Kelvin relation based on thermoelectrical potential}
\subsubsection{Seebeck effect in thermoelectric circuits}
The thermoelectrical potential \textit{ST} in the same thermoelectric material is a f state quantity. When a current is passed through a thermoelectric element, a thermoelectrical energy flow, i.e., $\mathit{SIT}=\mathit{IV}$, exists at each point of the element. The thermoelectrical energy flow decreases as it flows along the element from the high-temperature end to the low-temperature one. The reduced portion is used to convert to an electrical energy flow \textit{IV}. The conversion between the thermoelectrical energy flow and electrical energy flow can also be considered as the conversion of the thermal energy flow to electrical energy flow during the Seebeck effect, because the thermoelectrical energy flow is derived from the thermal energy flow.
\subsubsection{Peltier effect in thermoelectric circuit}
\begin{figure}[b]
	\includegraphics[width=8.6cm]{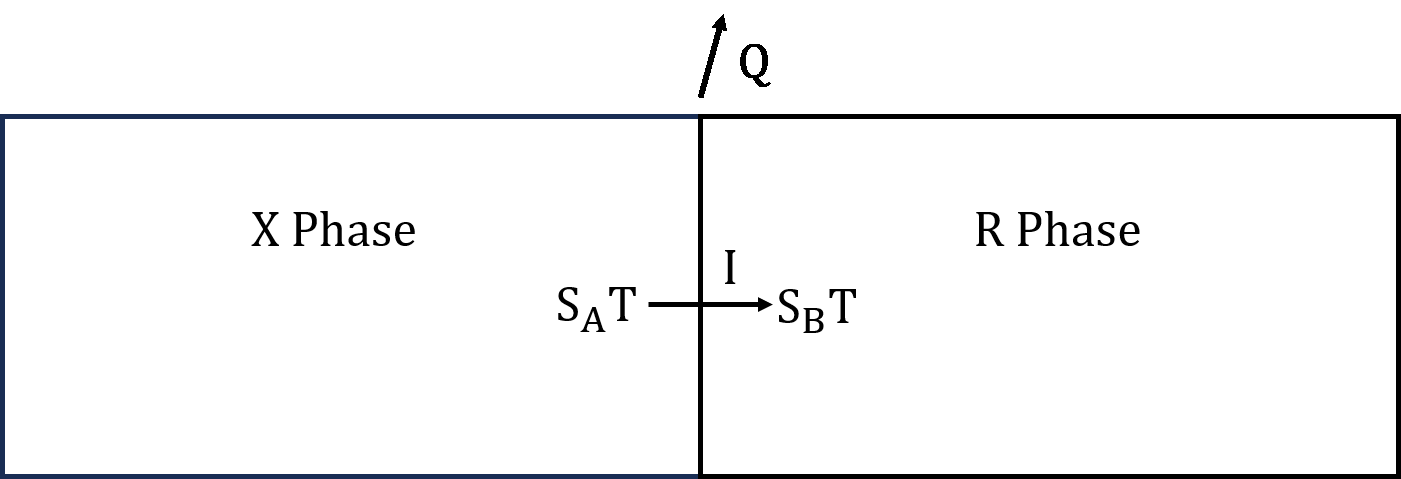}
	\caption{\label{fig:2}Diagram of the thermoelectrical potential at the interface of different materials.}
\end{figure}

According to the definition of thermoelectrical potential, at the same temperature, the thermoelectrical potentials of different thermoelectric materials are different, because of their distinct Seebeck coefficients, i.e., $(\mathit{S}_{A}-\mathit{S}_{B})\mathit{T}=\mathit{S}_{A}\mathit{T}-\mathit{S}_{B}\mathit{T}$. Therefore, the thermoelectrical potentials at the interface of two materials are discontinuous, as shown in Fig.\ref{fig:2}. The thermoelectrical energy flow $(\mathit{S}_{A}-\mathit{S}_{B})\mathit{TI}$ is also discontinuous when a current is passed. This phenomenon indicates that the circuit needs to have an exchange of thermoelectrical energy flow with the outside region to maintain a stable operation of its own circuit. 
\subsubsection{Kelvin relation}
The Peltier coefficient is defined as $\mathit{\Pi}=\mathit{Q}/\mathit{I}$, where \textit{Q} is the flow of thermal energy absorbed or released from the circuit at the interface, and \textit{I} is the current passing through the interface. The effect of the thermal energy flow is not to generate electrical energy flow but to compensate for the discontinuity in the thermoelectrical energy flow at the interface. Thus, the law of conservation of energy is obtained, i.e.,
\begin{eqnarray}
	\mathit{Q}=(\mathit{S}_{A}-\mathit{S}_{B})\mathit{IT}
\end{eqnarray}
This equation is not only for the conservation of energy but also for the conversion of energy in different forms. Although both ends of the equation are energy dimensions, \textit{Q} is the energy flow dimension of heat (joules per second), and the thermoelectric energy flow \textit{SIT} is the energy flow dimension of electricity (watts). Substituting the equation for the Peltier coefficient provides the Kelvin relation,
\begin{eqnarray}
	\mathit{\Pi}=\dfrac{(\mathit{S}_{A}-\mathit{S}_{B})\mathit{IT}}{\mathit{I}}=(\mathit{S}_{A}-\mathit{S}_{B})\mathit{T}
\end{eqnarray}
Thus, the Peltier coefficient is not only the ratio of the thermal energy flow \textit{Q} to the current \textit{I} at the interface but is also the thermoelectrical potential difference between the two materials. These characteristics further elucidate the physical significance of the Peltier coefficient. The energy conversion equation demonstrates that the Peltier effects is the mutual conversion of thermal and thermoelectrical energies, rather than the mutual conversion of thermal and electrical energies as reported in the literature.
\section{\label{sec:level8}CONCLUSIONS}
(1)	In this paper, the physical mechanism of Seebeck effect is analysed. The result shows the existence of a new potential field \textit{ST} in the thermoelectric material. This field can exert a force on the charge carriers in the thermoelectric material and may drive them to undergo a directional motion. This new potential field is named as the thermoelectrical potential. In an open-circuit state, the electrostatic potential can be balanced with the thermoelectrical potential in the material, ensuring that the total internal current is zero. In a closed-circuit state, the electric energy flow comes from the conversion of thermoelectric energy flow. The conversion can also be considered as the conversion between thermal energy and electrical energy flows, because the thermoelectrical energy flow is derived from the thermal energy flow.
\par
(2)	The analysis of the physical mechanism of the Peltier effect further confirms the proposed potential field. Without considering the resistance of the interface, the voltages on both sides of the interface are equal, and the electrical energy flow remains constant as the current passes through the interface. Therefore, another type of energy flow difference, namely, the thermoelectrical energy flow difference should exist. The Peltier effect at the interface of different thermoelectrical materials is not the conversion between thermal energy flow and electrical energy flow, but the conversion between thermal energy flow and thermoelectric energy flow.
\par
(3)	The Kelvin relation has been rederived using the introduced thermoelectrical potential. The results show that the proposed thermoelectrical potential elucidates the physical image of the thermoelectric conversion, and the derivation process can be simplified.
\begin{acknowledgments}
This work is supported by National Natural Science Foundation of China (No. 52250273).
\end{acknowledgments}
\bibliographystyle{unsrt}
\bibliography{references}
\end{document}